# Liquid general anesthetics lower critical temperatures in plasma membrane vesicles

**Ellyn Gray[†], Joshua Karslake[†], Benjamin B. Machta[‡]*, Sarah L. Veatch[†]***

[†] Department of Biophysics, University of Michigan, Ann Arbor MI 48109
[‡] Lewis-Sigler Institute for Integrative Genomics, Princeton University, Princeton NJ 08544

*co-corresponding authors

**ABSTRACT**

A large and diverse array of small hydrophobic molecules induce general anesthesia. Their efficacy as anesthetics has been shown to correlate both with their affinity for a hydrophobic environment and with their potency in inhibiting certain ligand gated ion channels. Here we explore the effects that n-alcohols and other liquid anesthetics have on the two-dimensional miscibility critical point observed in cell derived giant plasma membrane vesicles (GPMVs). We show that anesthetics depress the critical temperature ($T_c$) of these GPMVs without strongly altering the ratio of the two liquid phases found below $T_c$. The magnitude of this affect is consistent across n-alcohols when their concentration is rescaled by the median anesthetic concentration ($AC_{50}$) for tadpole anesthesia, but not when plotted against the overall concentration in solution. At $AC_{50}$ we see a 4°C downward shift in $T_c$, much larger than is typically seen in the main chain transition at these anesthetic concentrations. GPMV miscibility critical temperatures are also lowered to a similar extent by propofol, phenylethanol, and isopropanol when added at anesthetic concentrations, but not by tetradecanol or 2,6 diterbutylphenol, two structural analogs of general anesthetics that are hydrophobic but have no anesthetic potency. We propose that liquid general anesthetics provide an experimental tool for lowering critical temperatures in plasma membranes of intact cells, which we predict will reduce lipid-mediated heterogeneity in a way that is complimentary to increasing or decreasing cholesterol. Also, several possible implications of our results are discussed in the context of current models of anesthetic action on ligand gated ion channels.

**INTRODUCTION**

A large number of small molecules induce clinically similar general anesthesia, ranging from the noble gas Xenon to larger organic molecules. Since the early 20[th] century it has been known that the potency of a general anesthetic is roughly proportional to its oil:water partition coefficient over more than five orders of magnitude in overall concentration (3). This striking correlation along with the structural diversity of general anesthetics has led many to speculate that anesthesia is induced through non-specific effects on the physical properties of membrane lipids, and not through specific interactions with proteins. It has been proposed that this could be accomplished by altering the pressure profile of the membrane (4), the lateral organization of the membrane (5, 6), or the mechanical properties of axonal membranes (7). In support of these theories, general anesthetics were shown to decrease lipid chain ordering (8), increase membrane fluidity (9) and lower the transition temperature into a gel phase (10, 11).

Strong arguments have been made against membrane-mediated models of general anesthesia over the past several decades. First, although anesthetics do partition into membranes roughly in proportion to their potency (12), there are a variety of hydrophobic small molecules that partition strongly into membranes but have reduced or no anesthetic potency, some of which are structural analogs of well characterized anesthetics (8, 13-15). Second, it has been argued that the effects that anesthetics have on the physical properties of membranes are too small to be functionally significant at their anesthetic dose, as many properties are easily mimicked by raising temperature <1°C (16).

Instead, recent attention has focused on protein models.  It is now well known that the function of many ligand gated ion channels are sensitive to the presence of anesthetics (17) and it is widely (though not universally, see for example (7, 18, 19)) held that the anesthetic response is primarily due to the altered functioning of these ligand-gated channels.  Recent experiments have shown that this sensitivity can be modulated by mutating specific amino acid residues (20).  A recent crystallographic study has localized anesthetics in the vicinity of a proposed binding site of GLIC, a prokaryotic ligand gated ion channel that is sensitive to anesthetics, although the resulting structure most closely resembles the anesthetic destabilized 'open' state (21).  These and other related results are widely interpreted to imply that anesthetics exert their influence on channels by binding to specific sites on target molecules.

In parallel, the past decade has clarified the functional roles of lipids in biological processes at the plasma membrane of animal cells.  Relevant to the current study, it is now believed that proteins are not uniformly distributed in mammalian cell membranes, but that lipids help to laterally organize proteins into correlated structures with dimensions ranging between 10 and 100nm often termed lipid rafts or lipid shells (22-24).  These structures likely arise, at least in part, because plasma membranes can separate into two equilibrium liquid phases at low temperatures, called liquid-ordered ($L_o$) and liquid-disordered ($L_d$) (25).  We have recently proposed a model of plasma membrane heterogeneity that attributes this structure to the presence of a miscibility critical point in cell plasma membranes below growth temperatures (1, 26-28). This model is supported by experimental observations of critical behavior in giant plasma membrane vesicles (GPMVs) isolated from living cells (29, 30).  This shares essential elements of earlier models that focus on fluctuations in single component bilayer membranes in the vicinity of the gel-liquid transition (26, 27) or in single or multicomponent monolayer membranes near a miscibility critical point (28, 31, 32).  All of these models predict that the size, composition, and life-time of lipid mediated structures in intact cells at physiological temperature will be sensitive to the temperature of the underlying critical point.

In this study, we explore how liquid general anesthetics modulate the miscibility transition in plasma membrane vesicles isolated from RBL-2H3 cells.  This work is motivated by past experiments in purified model membranes that demonstrate the miscibility transition can be sensitive to small perturbations in composition (33-37).  We find that most liquid general anesthetics investigated lower critical temperatures in GPMVs by roughly 4°C at their anesthetic dose, and by larger amounts at higher concentrations.  This modulation is expected to reduce the magnitude of heterogeneity in plasma membranes at physiological temperature, and could in principle suppress the functions of proteins or protein networks that are sensitive to lipid-mediated structures.  There is some evidence that the localization and function of ligand gated ion channels, including some implicated in the anesthetic response are sensitive to 'raft' heterogeneity (38-42).  This suggests that anesthetics may act, at least in part, through their effects on membrane miscibility.

## MATERIALS AND METHODS

*Materials:*
Rat basophilic leukemia (RBL-2H3) cells were a gift from Barbara Baird and David Holowka. The fluorescent probe DiI-$C_{12}$ was purchased from Life Technologies (Carlsbad, CA).

Formaldehyde was obtained from Thermo Fisher Scientific (Waltham, MA). Diththiothreitol, ethanol, 1-propanol, 2-propanol, octanol, decanol, tetradecanol, propofol, phenylethanol, 2,6 ditertbutylphenol, dimethylsulfoxide (DMSO), and all additional reagents unless otherwise specified were purchased from Sigma Aldrich (St. Louis, MO).

*Cells and GPMV preparation:*
RBL-2H3 cells were grown and maintained with media containing MEM, 20% FBS, 10 μg/ml gentamicin sulfate in 25cm$^2$ flasks as described previously (43). Cells were labeled for 10min with 2μg/mL DiI-C$_{12}$ in methanol (1% final concentration) then GPMVs were prepared as described previously (25), by incubating cells in a buffer containing 2mM DTT, 25mM formaldehyde, 150mM NaCl, 2mM CaCl$_2$, and 20mM HEPES pH 7.4 at 37°C for 60min, after which they were isolated by separating buffer from adherent cells. GPMVs were treated for five minutes with anesthetics at the specified concentrations by diluting GPMV suspensions at most 1:1 into buffer containing anesthetics for at least 15 min prior to imaging. In control experiments (no anesthetic added) we observe no effect of diluting GPMV suspensions into additional buffer. Tetradecanol and 2,6 ditert-butylphenol were dissolved in DMSO prior to being added to GPMVs because these compounds have low water solubility. The maximum final concentration of DMSO was 14mM. Significantly higher DMSO concentrations (600mM) did not alter results in control experiments, and results were indistinguishable within error for GPMVs incubated with 11μM propofol in the presence or absence of 14mM DMSO.

*Fluorescence microscopy:*
GPMVs were imaged between two #1.5 coverslips sealed with a thin layer of Dow Corning high vacuum grease. Images were acquired using an inverted microscope (IX81; Olympus, Center Valley, PA) using a 40X 0.95 NA air objective, a Cy3 filter-set (Chroma Technology, Bellows Falls, VT) and an SCMOS camera (Neo; Andor, South Windsor, CT). The sample chamber was attached to a homebuilt temperature stage consisting of a peltier thermoelectric device, water-circulating heat sink (Custom Thermoelectric, Bishopville, MD), and a PID-type controller unit (Oven Industries, Mechanicsburg, PA). The sample was adhered using a thin layer of thermal grease to a copper plate in thermal contact with one side of the peltier device. Temperature was measured with a thermister probe mounted on the copper plate close to the sample.

*Transition temperature determination:*
Several (typically 5) fields containing numerous GPMVs (10-200) from the same sample were imaged at temperatures ranging from 25°C to 5°C in increments of 2 to 3°C. At each temperature, 200 to 500 GPMVs were imaged. The percent of GPMVs containing two coexisting liquid phases (%separated) was determined from images acquired at each temperature (T) using a custom program written in Matlab (MathWorks, Natick, MA) and errors are given by counting statistics. Critical temperatures (T$_C$) are obtained through a nonlinear least squares fit to the following sigmoidal curve:
$$\%seperated = 100 \times \left(1 - \frac{1}{1+e^{-(T-T_C)/B}}\right),$$
where B is a constant that describes the slope of the curve at the transition temperature. By this definition, T$_C$ corresponds to the temperature where fifty percent of vesicles contain coexisting liquid phases. Errors in absolute transition temperatures (σ$_T$) for a single measurement are extracted directly from the fit.

Absolute critical temperatures for GPMV preparations can show significant day-to-day variation, therefore we report critical temperature shifts in the presence of anesthetics compared to a control sample investigated from the same preparation of GPMVs. The error bounds for a given critical temperature shift ($\Delta T_C$) is given by $\sigma_{\Delta TC} = \sqrt{\sigma_{T0}^2 + \sigma_T^2}$, where $\sigma_{T0}$ is the error associated with determining the critical temperature of the control sample and $\sigma_T$ is the error associated in determining the critical temperature of GPMVs with anesthetic. In most cases, we report a critical temperature shift that is a weighted average of $\Delta T_C$ values determined from multiple experiments conducted on different days (typically 3-4). Weights are given by $w = \frac{1}{\sigma_i^2} / \sum \frac{1}{\sigma_i^2}$, where $\sigma_i$ is the error for each $\Delta T_C$ included in the average. Error bounds on average points ($\sigma$) are given by $\sigma = \sqrt{1 / \sum \frac{1}{\sigma_i^2}}$.

**RESULTS**

*Ethanol lowers critical temperatures in isolated plasma membrane vesicles.*

Giant plasma membrane vesicles (GPMVs) isolated from RBL-2H3 cells contain a single liquid phase at elevated temperatures and coexisting liquid-ordered ($L_o$) and liquid-disordered ($L_d$) phases at low temperatures. Phase separation is easily visualized by monitoring the lateral distribution of the fluorescent lipid analog DiI-$C_{12}$ in GPMVs using conventional fluorescence microscopy. While individual GPMVs have well defined transition temperatures, there is significant variation in transition temperatures between vesicles isolated from a single flask of cells. For this reason, we report the average transition temperature for a sample of GPMVs by determining the fraction of vesicles containing coexisting phases as a function of temperature, as shown in Fig.1. This distribution is fit to a sigmoid function to extract the temperature where 50% of vesicles have undergone a transition, which we report as the transition temperature of the sample. For the control sample in Fig. 1 *A*, we extract an average transition temperature of 15.9±0.3°C, where the error is extracted directly from the fit.

When GPMVs isolated from the same dish of cells are incubated with ethanol prior to imaging, we observe the average transition temperature at a lower value. For the case of 120mM ethanol shown in Fig. 1 *A*, we measure an average transition temperature of ethanol treated GPMVs to be 12.5±0.6°C, which represents a downwards shift of $\Delta T_C$ = -3.4±0.7°C compared to untreated vesicles. This shift in average transition temperature is most apparent when vesicles are visualized at a temperature where a finite fraction of vesicles are phase separated in untreated GPMVs. When ethanol treated vesicles are imaged at this same temperature, many fewer vesicles contain coexisting liquid phases. We also find that the magnitude of $\Delta T_C$ depends on the concentration of anesthetic incubated with GPMVs. Fig. 1 *B* shows the fraction of phase separated GPMVs as a function of temperature for control GPMVs and GPMVs incubated with 120mM, 240mM, or 600mM ethanol prior to imaging, with all GPMVs prepared from the same flask of cells. Extracted transition temperatures are shown in Fig. 1 *C* and are monotonically decreasing with increasing ethanol concentration, with nonlinearities at low concentrations.

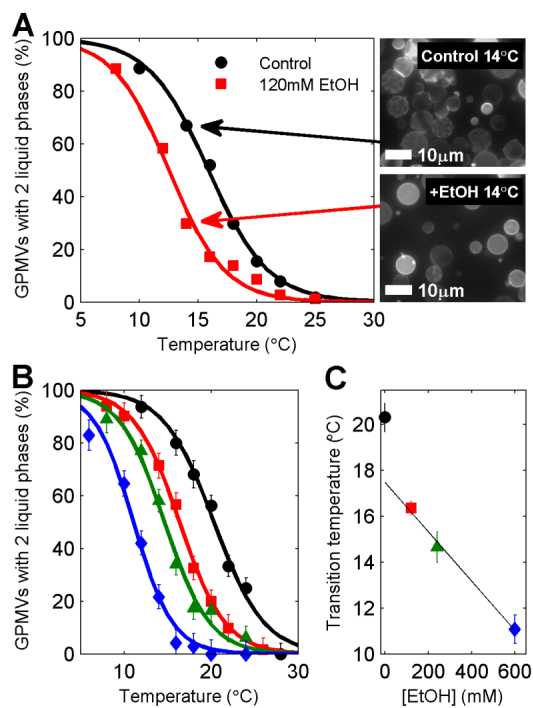

**FIG. 1: Ethanol lowers transition temperatures in GPMVs.** *(A)* Untreated GPMVs isolated from a single flask of cells show a broad distribution of transition temperatures. We quantify average transition temperatures by measuring the fraction of GPMVs with coexisting phases at multiple temperatures *(black points)* and fit this distribution to extract the temperature where 50% of GPMVs contain coexisting liquid phases. When ethanol (EtOH) is added to vesicles from the same flask of cells prior to imaging, the distribution of phase separated vesicles shifts to a lower temperature *(red points)*, indicating that the transition temperature is lower. This downward shift in transition temperature is most apparent when comparing fields of vesicles imaged at 14°C in this experiment, where the majority of control GPMVs contain coexisting liquid phases whereas most ethanol treated vesicles are uniform. *(B, C)* The transition temperature shift is dependent on the concentration of ethanol incubated with GPMVs. The curves in *B* are fit to obtain the average transition temperature reported in *C* and are drawn with the same symbols.

In addition to variation of transition temperatures between vesicles in a given GPMV preparation, we also observe significant day to day variation when average transition temperatures are measured. This is demonstrated by comparing the control curves in Fig. 1, *A* and *B*, which yield different absolute transition temperatures (15.9±0.3°C vs. 20.3±0.6°C). Even so, we find that the magnitude of the downward shift in transition temperatures observed after the addition of liquid general anesthetics is robust. For the example of incubation with 120mM ethanol, we observe an average transition temperature shift of $\Delta T_C$ = -3.5±0.6°C when averaged over 3 distinct experiments, where the error is the standard deviation between the three measurements.

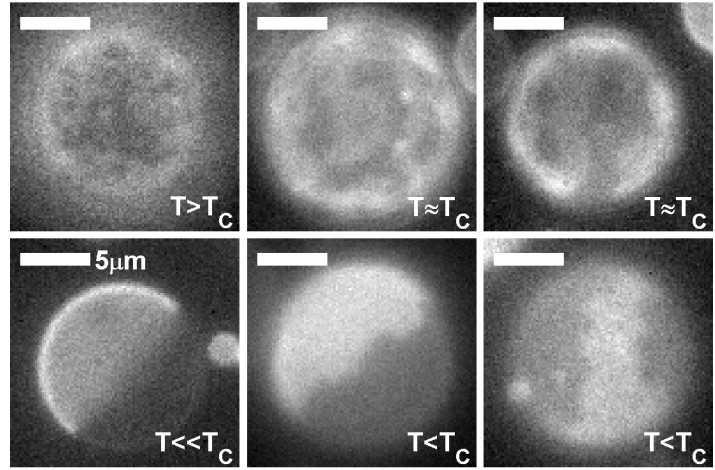

**FIG. 2: Ethanol treated vesicles retain critical fluctuations.** Representative images of different GPMVs imaged in 120mM ethanol demonstrating a range of critical phenomena including dynamic super-critical fluctuations for $T \geq T_C$ and phase boundary fluctuations for $T \leq T_C$. At low temperatures ($T \ll T_C$), vesicles separate into roughly equal surface fractions of liquid-ordered and liquid-disordered phases. Vesicle imaged over a range of temperatures between 11°C and 16°C.

*Ethanol treated vesicles retain critical fluctuations*

In addition to undergoing a miscibility transition near room temperature, untreated GPMVs exhibit robust critical fluctuations within several degrees of their miscibility transition temperature. The size and contrast of fluctuations in both synthetic and isolated biological membranes have been previously shown to be consistent with belonging to the 2D Ising model universality class (29, 44). Hallmarks of criticality in membranes include micron-sized and dynamic composition fluctuations within ~1°C above the miscibility transition temperature and undulating phase boundaries within several degrees below the miscibility transition temperature (44, 45). Vesicles with critical compositions also contain roughly equal surface fractions of phases at temperatures just below the phase transition. Untreated vesicles show weak temperature dependence in the surface fraction of coexisting phases, consistent with their being a nearly vertical rectilinear diameter (1).

While incubation with different concentrations of ethanol lowers transition temperatures in GPMVs, hallmarks of critical behavior are still observed in these vesicles, albeit at a lower temperature, as shown in Fig. 2. Ethanol treated vesicles can be found to contain large and dynamic composition fluctuations above their transition temperature, undulating phase boundaries below their transition temperature, and roughly equal surface fractions of phases in vesicles well below their phase transition temperature. These results suggest that although membrane composition is altered through incubation with ethanol, this composition change acts to shift the critical temperature without shifting vesicle composition away from a critical composition

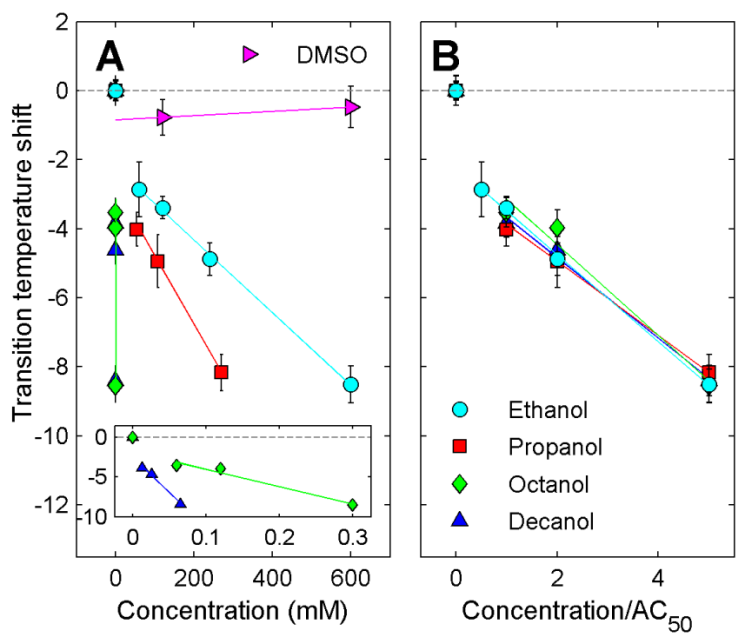

**FIG. 3: Critical temperature shift in GPMVs scales with anesthetic potency.**
*(A)* Measured critical temperature shifts in GPMVs incubated with a series of n-alcohol general anesthetics as a function of aqueous concentration of n-alcohol. Data points represent a weighted average over multiple experiments, as described in Methods and error bounds are propagated through this average. Lines are fits to all points excluding 0mM n-alcohol. *(B)* Data from *A* repotted with concentration scaled by the published anesthetic dose ($AC_{50}$) of these compounds (2).

*Liquid alcohol anesthetics lower critical temperatures by 4°C at their anesthetic dose.*

Ethanol is a member of a series of n-alcohol general anesthetics, and we have also investigated how other compounds in this series modulate phase transition temperatures and fluctuations in isolated GPMVs. As summarized in Fig. 3 *A*, all n-alcohol anesthetics investigated lower critical temperatures in GPMVs compared to untreated vesicles isolated from the same flask of cells although the concentration required depends strongly on the alcohol used. Longer chain alcohols such as octanol or decanol are the most potent. Similar to ethanol, GPMVs treated with longer chain n-alcohols also retain hallmarks of critical behavior near and below their transition temperature. As a control, we investigated GPMVs incubated with DMSO, where we observe no shift in transition temperatures when DMSO is included at 600mM.

Remarkably, plots of $\Delta T_C$ vs. n-alcohol concentration collapse onto a single curve within error when n-alcohol concentration is re-scaled by their anesthetic potency, as shown in Fig. 3 *B*. The anesthetic dose, or $AC_{50}$ value, is defined as the anesthetic concentration where 50% of tadpoles lose their righting reflex, as reported in a previous study (2). $AC_{50}$ values for the n-alcohols investigated are 120mM (ethanol), 54mM (isopropanol), 60μM (octanol), and 13μM (decanol). For all compounds, we observe a downward shift of 4°C when added at their anesthetic dose. We note that this is significantly larger than the ~1°C downward shift observed in the main chain transition temperature of DPPC with the same concentration of n-alcohol anesthetic (16).

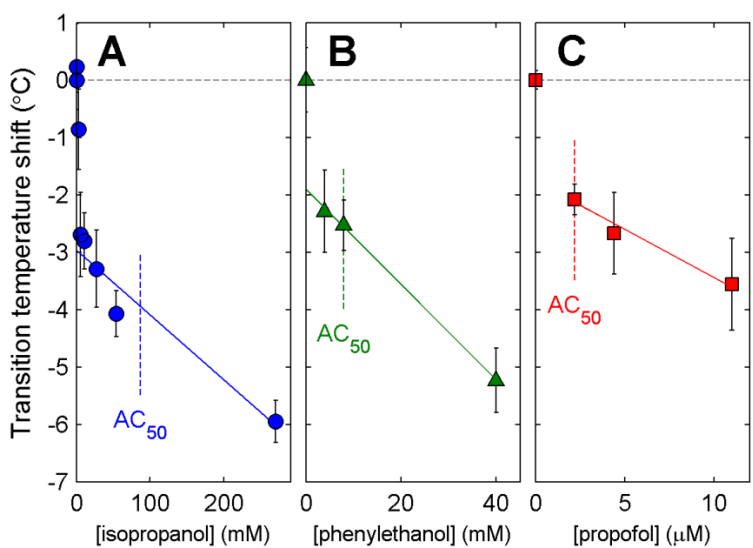

**FIG. 4: Non n-alcohol liquid anesthetics lower critical temperatures at their anesthetic dose.** Data points represent a weighted average over at least two separate experiments and error bounds are propagated through this average, as described in Methods. Lines are fits to all points excluding 0mM for phenylethanol and propofol and <50µM for isopropanol. The estimated anesthetic doses for these compounds are indicated by dashed lines and are determined as described in the main text.

*Non n-alcohol anesthetics also lower critical temperatures at their anesthetic dose.*

In order to explore if our observations regarding n-alcohols could be generalized to other liquid anesthetics, we also investigated the impacts of three additional anesthetic compounds in Fig. 4. Isopropanol is an isomer of 1-propanol, and has a slightly reduced concentration required for blocking power on motor fibers from sciatic nerves of frogs (351mM vs. 218mM) (46). If we re-scale this value using the $AC_{50}$ value for 1-propanol to influence the righting reflex of tadpoles, we estimate an $AC_{50}$ value for isopropanol to be 87mM, as indicated on Fig. 4 *A*. At this concentration, the critical temperature is depressed by ~4°C, as is seen for n-alcohol anesthetics. Phenylethanol (PEtOH) is a local anesthetic which has been previously demonstrated to alter helix aggregation in bacterial membranes and single component model membranes (47). Again, we observe that adding PEtOH to GPMVs acts to reduce critical temperatures, with a 4°C downward shift at approximately 25mM. At a possibly more clinically relevant concentration of 8mM (47), we observe a downward shift of 2.5±0.4°C. Finally, we explored the intravenous general anesthetic propofol, and again find that propofol lowers critical temperatures when added to GPMVs. Previous work has measured an $AC_{50}$ value of between 2.2 and 2.5µM (48), and at this concentration we observe a 2.1±0.3°C downward shift in GPMV critical temperature. For all three compounds, we observe a linear concentration dependence with a negative y-intercept of about -2°C as was observed with other n-alcohols, indicating nonlinearity at low concentrations. Overall, we find that these compounds reveal the same pattern observed for n-alcohol anesthetics, although in some cases the magnitude is reduced.

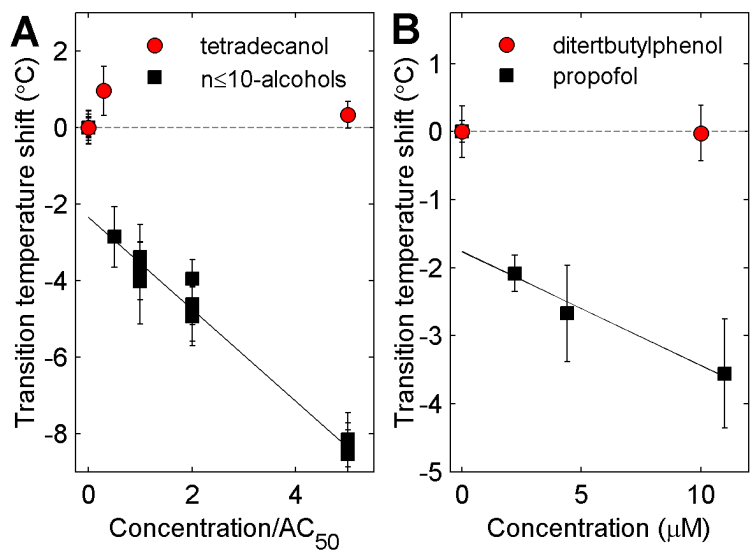

**FIG. 5 Structurally similar non-anesthetics do not lower critical temperatures.**
*(A)* Tetradecanol is an n-alcohol but is not active as a general anesthetic. Tetradecanol does not lower critical temperatures when added to GPMVs. n-alcohol data for n≤10 is re-plotted from Fig. 3 *C*. *(B)* 2,6 diterbutylphenol is structurally similar to propofol but contains two additional methyl groups and is not active as a general anesthetic. 2,6 diterbutylphenol does not lower critical temperatures in GPMVs. Propofol data re-plotted from Fig. 4 *C*.

*Membrane soluble molecules that are not anesthetics do not lower critical temperatures.*

There are numerous hydrophobic small molecules that are structurally similar to general anesthetics, but do not exhibit any general anesthetic activity. One well-studied example is long chain n-alcohols with n≥12. This so-called 'anesthetic cutoff effect' has been used to argue against a lipid-mediated mechanism of general anesthesia (e.g. (13, 15)). Fig. 5 *A* shows that tetradecanol (n=14) does not alter critical temperatures in GPMVs even when added at concentrations up to 5μM, which is the measured $AC_{50}$ value for dodecanol (n=12). A second example of a hydrophobic small molecule without anesthetic activity is 2,6 diterbutylphenol (15) and is shown in Fig. 5 *B*. 2,6 diterbutylphenol is structurally similar to propofol but contains two additional methyl groups and is therefore more hydrophobic. Similar to tetradecanol, we again observe no shift in critical temperatures when 2,6 diterbutylphenol is added to suspensions of GPMVs, even at concentrations where propofol decreases $T_C$ by roughly 4°C.

## DISCUSSION

Our observation that liquid general anesthetics lower critical temperatures in isolated GPMVs suggest that these compounds could be a useful experimental tool for probing the effects of lipid-mediated heterogeneity in a range of cellular processes. Within the critical fluctuation model of plasma membrane heterogeneity, a compound that acts to lower $T_C$ would also reduce the size, lifetime, and compositional contrast of lipid-mediated domains in the plasma membrane at physiological temperature. In this context, liquid general anesthetics are expected to act as inhibitors of lipid heterogeneity, and should interfere with the functioning of proteins or protein networks that exploit this type of lipid structure to perform their cellular functions. Unlike cholesterol reduction, which modulates the miscibility transition in GPMVs by reducing the surface fraction of the liquid-ordered phase (30, 49), liquid general anesthetics alter membrane mixing properties while maintaining membranes at a critical composition. We expect anesthetics to be useful as a complimentary membrane perturbation to the commonly used cholesterol modulation, by perturbing lipid mediated heterogeneity through an alternate mechanism (Fig. 6). Anesthetics may produce results that are easier to interpret as lipid mediated interactions will be reduced without changing the surface area available to compounds that have a strong preference for either the liquid-ordered or liquid-disordered phase. In addition, liquid general anesthetics alter membrane mixing properties without the large changes in global membrane composition that accompany cholesterol reduction, since cholesterol typically makes up ~40 mol% of the intact plasma membrane (50).

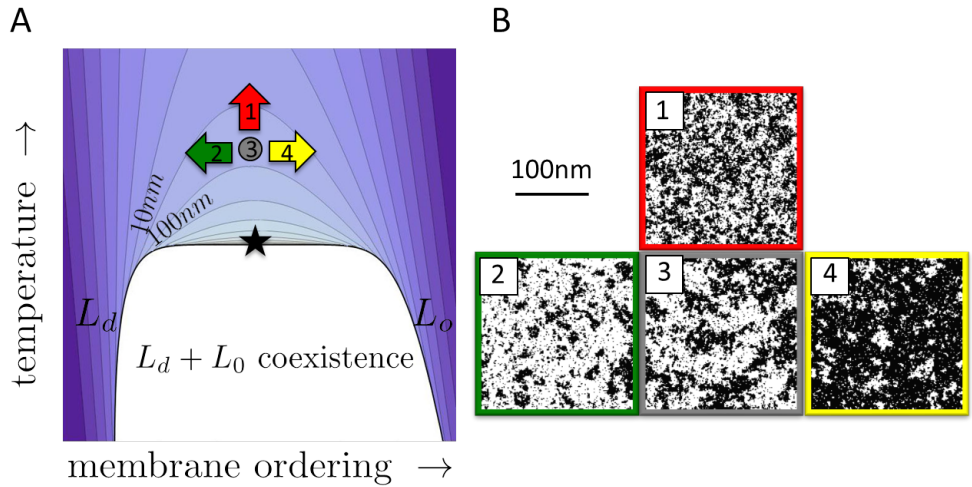

**FIG. 6:** *(A)* Proposed schematic phase diagram of a cell membrane. Normal physiological conditions *(grey circle)* reside above a miscibility critical point *(star)* below which the membrane separates into coexisting $L_o$ and $L_d$ phases. Contours are of constant correlation length as described previously (1) and physiological conditions predicted to have a correlation length of roughly 20nm. At constant physiological temperature, addition of general anesthetics is expected to shift membranes in the direction of the *red arrow (1)*, increasing the temperature difference from the critical point. The more familiar perturbations of cholesterol depletion (loading) act to decrease (increase) membrane order, acting like the *green* (*red*) *arrows* on the phase diagram marked *2* and *4*. *(B)* Perturbations affect the lateral structure of the membrane as depicted schematically with snapshots from a lattice Ising model, which shares broad features of its phase diagram. Under physiological conditions (*3*), there is no macroscopic phase separation, but super-critical fluctuations lead to structure that is much larger than the size of microscopic components, and there is roughly an equal area of ordered and disordered components. The common perturbations of cholesterol depletion (*2*) and loading (*4*) primarily act to increase the surface fraction of disordered (*white*) or ordered (*black*) membrane respectively. In contrast, liquid general anesthetics perturb membrane structure, but do not affect the ordered to disordered ratio (*1*). Instead they reduce the size, lifetime and contrast of fluctuations.

In the context of general anesthesia, our findings contribute to the current discussion regarding the molecular mechanisms through which liquid general anesthetics act on ligand gated ion channels. Within a model where general anesthetics act primarily through direct binding to ligand gated ion channels, our results indicate that these ligands also act to lower critical temperatures in membranes. More generally, we can infer that liquid general anesthetics partition to the interface between ordered and disordered liquid domains because these compounds lower the miscibility transition temperature without changing the surface fraction of low temperature phases (51). If ligand gated ion channels also localize to the interface, this could be a means to increase the local concentration of ligands close to the binding site, as has been noted previously (5). Alternately, a channel could alter its partitioning through the binding of anesthetic ligands if these compounds retain their phase preference upon binding to target proteins. Our observations using synthetic liquid anesthetics also suggest that endogenous ligands for these sites may also share the property that they lower critical temperatures in plasma membranes and localize to domain boundaries. It is possible that putative anesthetic sites provide a means of coupling membrane properties to the functioning of proteins.

We find it striking that the $T_C$ lowering effect scales remarkably well with anesthetic potency for the four n-alcohol general anesthetics investigated. To first order, this is the expected result because n-alcohols become more hydrophobic as n is increased and therefore partition more strongly into membranes (12). However, this striking correlation extends beyond the n-alcohol anesthetics to include tetradecanol, an n-alcohol which is expected to strongly partition into membranes but has no anesthetic potency. We observe no downward shift in $T_C$ in membranes containing tetradecanol, in good agreement with previous ESR studies which demonstrated that tetradecanol does not alter the ordering of natural membranes even at much higher membrane concentrations (8). $T_C$ is also not altered when the propofol analog 2,6 diterbutylphenol is added to GPMVs, again consistent with it lacking anesthetic activity. For the molecules investigated here, the shift in $T_C$ more accurately predicts general anesthetic potency than does membrane partitioning.

We observe a much larger shift in $T_c$ than is expected from a naïve extrapolation of results in other membrane assays, which typically show changes that can be mimicked by very small (<1K) changes in temperature (16). The magnitude of the anesthetic effect on the miscibility transition in GPMVs also highlights the differences between the membrane miscibility transition examined here and the main chain melting transition present in single-component membranes, where transition temperatures are reduced in the presence of anesthetic through freezing point depression. Previous work has demonstrated that miscibility transitions are generally more sensitive to perturbations in the form of impurities (33-36, 52). The miscibility transition in GPMVs is also different than the main chain transition in purified membranes in that it is a continuous, second order transition, tuned very close to its critical point (Fig. 6). The miscibility critical point is distinguished by an increased sensitivity of physical properties to changes in its reduced temperature and composition. Thus, a small shift in $T_c$ could lead to large changes in physical properties at a fixed temperature.

Our results hint at the intriguing possibility that anesthetics may exert some of their influence on proteins through their effects on the membrane's miscibility behavior. Plasma membrane composition has been implicated in controlling the spatial localization of membrane receptors and channels (38), and there is evidence that many channels including those implicated in the anesthetic response are localized in this fashion, including the gamma aminobutyric acid A (GABA-A) receptor (41) and nicotinic acetylcholine receptor (nAchR) (53). It is also possible that localization is tied in some way to function, either by providing a local lipid environment that either promotes or suppresses channel activation, by bringing receptors into close proximity with regulatory components, or by tuning the interactions between components of a regulatory network (see (54) and Fig. 6). Along these lines, it has been shown that plasma membrane cholesterol levels modulate the functioning many ion channels (38), including both GABA-A (42) and nAchR (40), although these changes could be mediated by direct binding to cholesterol or by specific interactions with neurosteroid modulators (55).

In principle, an ion channel regulated through its preferences for a local lipid environment could have its gating properties changed by any perturbation that modulates the mixing properties of lipids, even without specific binding of these compounds to the channels themselves. This includes liquid general anesthetics, which we demonstrate in this study to lower critical temperatures in isolated plasma membranes. Further experimental and modeling work is required to clarify if this type of regulation plays an important role in the specific case of functioning of the ligand gated ion channels responsible for general anesthesia, and if the 4°C shift observed in our study is adequate to significantly impact receptor functions. More generally, we expect that liquid general anesthetics will interfere with the large number of biological processes that have already been postulated to make use of membrane 'raft' heterogeneity to compartmentalize and organize their function.

## ACKNOWLEDGEMENTS

We thank Jing Wu for assistance with experiments and Keith Miller, James Sethna, and Dave Holowka for helpful conversations. BBM acknowledges partial support from the NIH (T32GM008267) and the Lewis-Sigler Fellowship program at Princeton University. Research was supported by the NIH:R00GM087810 (SLV) and startup funds from the University of Michigan.